\documentclass[conference]{IEEEtran}
\usepackage{cite}

\usepackage{amsmath,amssymb,amsfonts}
\usepackage{times}
\usepackage[ruled,vlined]{algorithm2e}
\usepackage{graphicx}
\usepackage{textcomp}
\usepackage{xcolor}
\def\BibTeX{{\rm B\kern-.05em{\sc i\kern-.025em b}\kern-.08em
    T\kern-.1667em\lower.7ex\hbox{E}\kern-.125emX}}
\usepackage{subfigure}
   
\usepackage{tikz,xcolor,hyperref}
\definecolor{lime}{HTML}{A6CE39}
\DeclareRobustCommand{\orcidicon}{%
\begin{tikzpicture}
\draw[lime, fill=lime] (0,0) 
circle [radius=0.16] 
node[white] {{\fontfamily{qag}\selectfont \tiny ID}};
\draw[white, fill=white] (-0.0625,0.095) 
circle [radius=0.007];
\end{tikzpicture}
\hspace{-2mm}
}
\foreach \x in {A, ..., Z}{%
	\expandafter\xdef\csname orcid\x\endcsname
	{\noexpand\href{https://orcid.org/\csname orcidauthor\x\endcsname}
	{\noexpand\orcidicon}}
}

\begin{document}

\title{High-Performance Mining of COVID-19 Open Research Datasets for Text Classification and Insights in Cloud Computing Environments}

\author{
\IEEEauthorblockN{Jie Zhao, Maria A. Rodriguez and Rajkumar Buyya}
\IEEEauthorblockA{\textit{Cloud Computing and Distributed Systems Laboratory} \\
\textit{School of Computing and Information Systems}\\
The University of Melbourne, Australia \\
Email: zhao.j4@student.unimelb.edu.au, \{maria.read, rbuyya\}@unimelb.edu.au
}
}
\maketitle

\begin{abstract}
COVID-19 global pandemic is an unprecedented health crisis. Since the outbreak, many researchers around the world have produced an extensive collection of literatures. For the research community and the general public to digest, it is crucial to analyse the text and provide insights in a timely manner, which requires a considerable amount of computational power. Clouding computing has been widely adopted in academia and industry in recent years. In particular, hybrid cloud is gaining popularity since its two-fold benefits: utilising existing resource to save cost and using additional cloud service providers to gain assess to extra computing resources on demand. In this paper, we developed a system utilising the Aneka PaaS middleware with parallel processing and multi-cloud capability to accelerate the ETL and article categorising process using machine learning technology on a hybrid cloud. The result is then persisted for further referencing, searching and visualising. Our performance evaluation shows that the system can help with reducing processing time and achieving linear scalability. Beyond COVID-19, the application might be used directly in broader scholarly article indexing and analysing.
\end{abstract}

\section{Introduction}
COVID-19 is a global scale health crisis. Since the outbreak, a massive amount of research efforts have been poured into many aspects of this highly infectious disease. To help the research community, in March 2020, the White House and the Allen Institute for AI teamed up with many researchers and released the COVID-19 Open Research Dataset(CORD-19)\cite{wang-lo-2020-cord19}. As of 27/Jul/2020, CORD-19 contains over 199,000 research papers and nearly half of them are open-access with full text available\cite{kaggle}. The amount of articles' rapid growth has posed a challenge for the research community to keep up to date. Also, the general public is interested in many aspects of the disease, especially those findings related to day-to-day life. 
Hence, the CORD-19 dataset is shared on Kaggle. Community actions are required to help with developing tools that facilitate the understanding of virus\cite{kaggle} using machine learning(ML) technology. The ability to analyse the data and provide insights promptly pose a challenge since ETL and ML requires considerable computing power.

In recent years, both industry and academia have adopted cloud computing paradigm, especially in the form of Hybrid Cloud Environment (HCE).\cite{Calheiros2012} Adopting HCE enables user to utilise their existing computing infrastructure, and instantaneously acquires additional resources from external cloud service providers (CSP) on demand whenever requirement arises. In context of processing voluminous amounts of data such as the CORD-19 dataset, HCE provides cost saving measure by using an existing on-premise cloud, meanwhile, supplies the capability to gain extra computing, storage or networking resources from other CSP. However, building an application using HCE is also a demanding task; it requires detailed knowledge of cloud computing techniques.

In this context, we propose a system design and implementation to accelerate ETL process and text classification with ML technology in a hybrid cloud environment. The architecture design is aimed to address the following characteristics: scalability, availability, stability, high performance and portability. To achieve these goals and for ease of development, the proposed application deploys on top of Aneka Platform-as-a-Service (PaaS) platform.\cite{Vecchiola2009}  Aneka PaaS is a range of tools, providing high level Application Programming Interfaces(APIs) and Software Development Kits(SDKs) for simpleness of implementing a scalable application. It allows developers to focus on developing their program logic without spending too much time considering deployment and scalability. When additional resources are required, one can easily acquire extra resources from additional CSP via Aneka dynamic provisioning mechanism. 

In this paper, we make the following key contributions:
\begin{itemize}
	\item We present a system architecture design that achieves various characteristics: scalability, availability, stability, high performance and portability.
	\item The system is implemented and tested using real world CORD-19 dataset in a real hybrid cloud environment built using on-premise private cloud and the Melbourne Research Cloud(MRC).
	\item The architectural design can be easily generalized and quickly adopted in similar scenarios. The system is not only applicable for the CORD-19 dataset but also for a broader scholarly article indexing and analysing. 
\end{itemize}

The rest of the paper is organized as follows: Section \ref{sec:background} outlines the background information about Aneka PaaS and other technologies used in the system. System architecture is detailed and discussed in Section \ref{sec:arch}. Section \ref{sec:sysimpl} explains the system design and implementation by using UML diagrams and workflows, in addition, some example queries and visualizations are given. Afterwards, Section \ref{sec:eval} describes the testbed built on a hybrid cloud including an on-premise private cloud and MRC. Our performance evaluation shows linear scalability can be achieved by utilizing Aneka PaaS with little overheads. Finally, the last section concludes the paper with summary and future works.

\section{Background and Related Work}
\label{sec:background}
Since the CORD-19 dataset made available, it has been downloaded for more than 200,000 times and many applications have been created \cite{kaggle}. For instance, Amazon Web Service (AWS) provides a search engine over the CORD-19 dataset and a question answering system powered by AWS Kendra \cite{awscord19}. Azure \cite{azurecord19}, TekStack \cite{tekstack}, and COVID-Miner \cite{covidminer} developed full-text search engines. VIDAR-19, which extracts and visualizes risk factor from articles, was presented by F. Wolinski \cite{Wolinski2020}. COVID Seer \cite{covidseer} and COVID Explorer \cite{covidexplorer} were developed by the Pennsylvania State University. COVID Seer is a multi-facet search engine powered by ElasticSearch and COVID Explorer has visualization and advanced filtering features utilizing automatic unsupervised ML. 

One drawback with existing systems is that they are not keeping updated with the growing dataset. Some of these are still using the initial version of CORD-19, even though the current dataset has grown threefold since the initial release. Hence, we address this issue by introducing additional workflows to keep up-to-date with current version of CORD-19 dataset and ingest newly published articles. These workflows allow users to have a latest view of state-of-the-art COVID-19 research outputs.

To implement the system quickly, the best solution is to select some proven technologies in both industry and academia. After evaluating many difference technologies, we decide to go with the following: (a) Aneka PaaS \cite{Vecchiola2009} for core processing and deployment; (b) Microsoft .NET Framework and ML.net \cite{mldotnet} for development; (c) Grobid \cite{grobid} for ML based scientific paper parsing; (d) Minio \cite{minio} for scalable S3 compatible shared storage; (e) Elastic Stack \cite{elastic} including ElasticSearch for full-text indexing and Kibana for data visualization.

\begin{figure*}[h]
	\centerline{\includegraphics[width=0.8\textwidth]{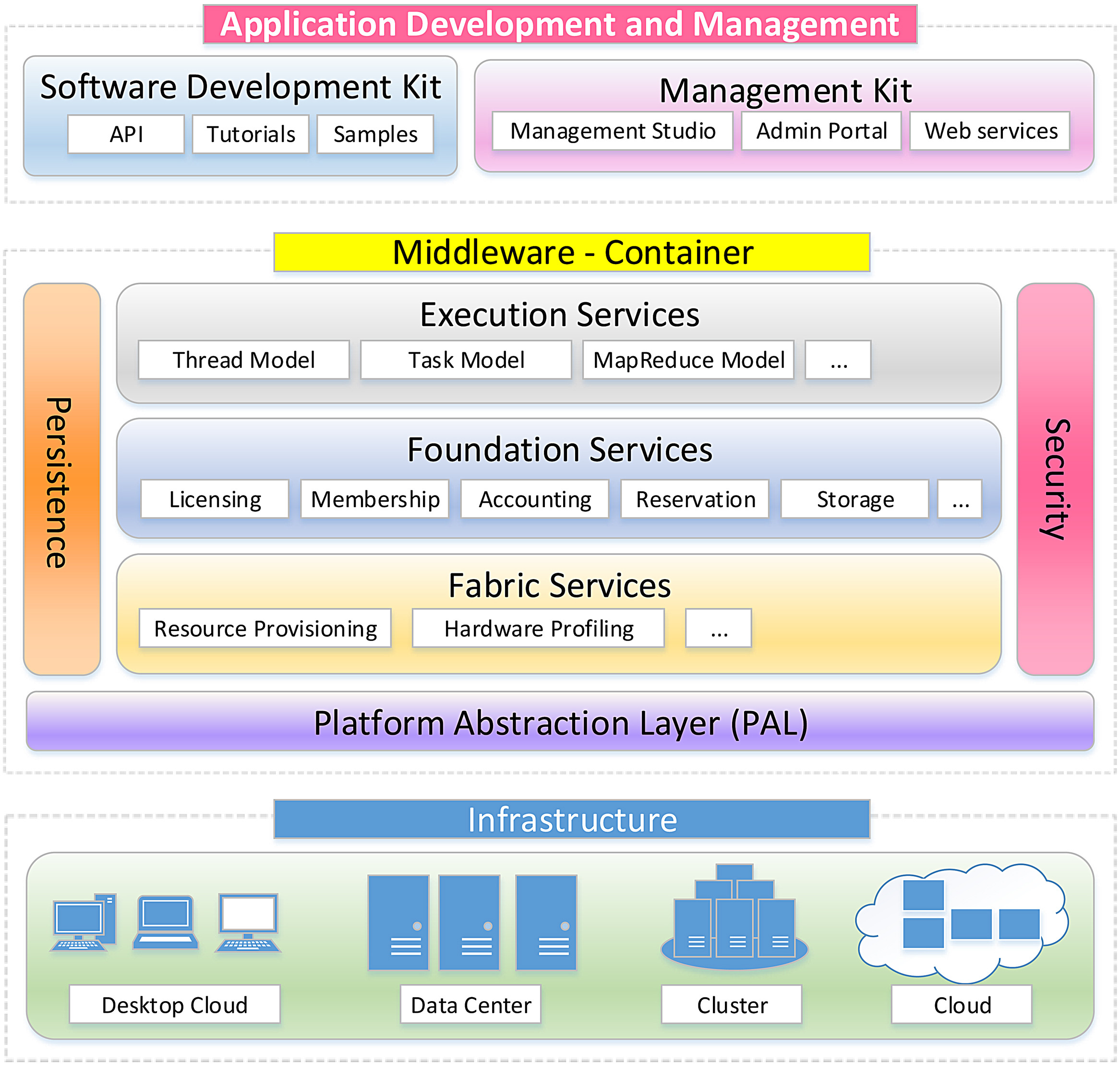}}
	\caption{Aneka Framework Overview\cite{NadjaranToosi2018} }
	\label{fig:aneka}
\end{figure*}

Aneka PaaS provides a platform for users and developers to develop and deploy distributed application with ease. An overview of Aneka is shown in Fig \ref{fig:aneka}. It comprises of three major layers with rich collection of components:
\begin{itemize}
	\item Application Development and Management Layer: This layer contains Software Development Kit (SDK) and management kit. SDK is a collection of Application Programming Interfaces (APIs), tutorials and examples to help users and developers getting started. Management kit includes Graphical User Interface (GUI) to assist with management. It comprises a management studio, admin portal and web services allowing users to view various status and statistics of the lower levels.
	\item Middleware Container Layer: This layer provides many services, such as execution services, foundation services , fabric services, etc. Among these, the execution services is the most crucial part for our application. It performs scheduling and execution with user's choice of QoS and strategy and supports four execution models: (1)Bag of Tasks (BoTs); (2)Distributed Threads (DTs); (3)MapReduce; and (4) Parameter Sweep Model (PSM). In our application, we want the processing to be completed as soon as possible. Therefore, we can use either BoTs or DTs model; we choose the default QoS strategy that uses all available computing resources.
	\item Infrastructure Layer: The bottom layer offers fundamental support to the above layers. In current version, Aneka allows to use resources with static provisioning in desktop cloud, data centre, cluster; and also dynamic provisioning in clouds such as Amazon EC2, Microsoft Azure, GoGrid,  etc. 
\end{itemize}

Microsoft .NET Framework/.NET Core is a free software development framework developed by Microsoft. It is a complete platform that supports various languages, such as C\#, VB, F\#, etc. With the strategic tradition to .NET Core, it also provides cross-platform support including Windows, Linux and MacOS. Since Aneka itself is developed with .NET framework, it becomes a natural choice for our application. In addition, Microsoft made an extensive machine learning framework for .NET developers. It implements many traditional and proven ML algorithms; users who don't have detailed knowledge of ML techniques can still easily use ML technology in their application. This significantly lowers the barrier for developer to utilize ML technology.

Elastic Stack (ES) \cite{elastic} is widely used in industry; it is also know as ELK Stack (ElasticSearch, Logstash, Kibana). Elastic Search is a distributed full-text indexing and search engine based on Apache Lucene library; Logstash provides data collection and log-parsing engine through various of agents called Beats; Kibaba is a data visualization platform using searching, filtering and aggregation functions provided by ElasticSearch. Since we implement customized data collections and processing logics by Aneka, we only use ElasticSearch and Kibana in our system. 

In addition to the technologies above, we also used an open source program called Grobid \cite{grobid} in our update workflow. Grobid is a machine learning library designed explicitly for extracting text data from technical or scientific documents. It is capable of converting PDF file to TEI/XML while maintaining section and structure format. It has been actively developed since 2008 and open-sourced in 2011.

\section{System Design and Architecture}\label{sec:arch}
\begin{figure*}[h]
	\centering
	{\includegraphics[width=0.9\textwidth]{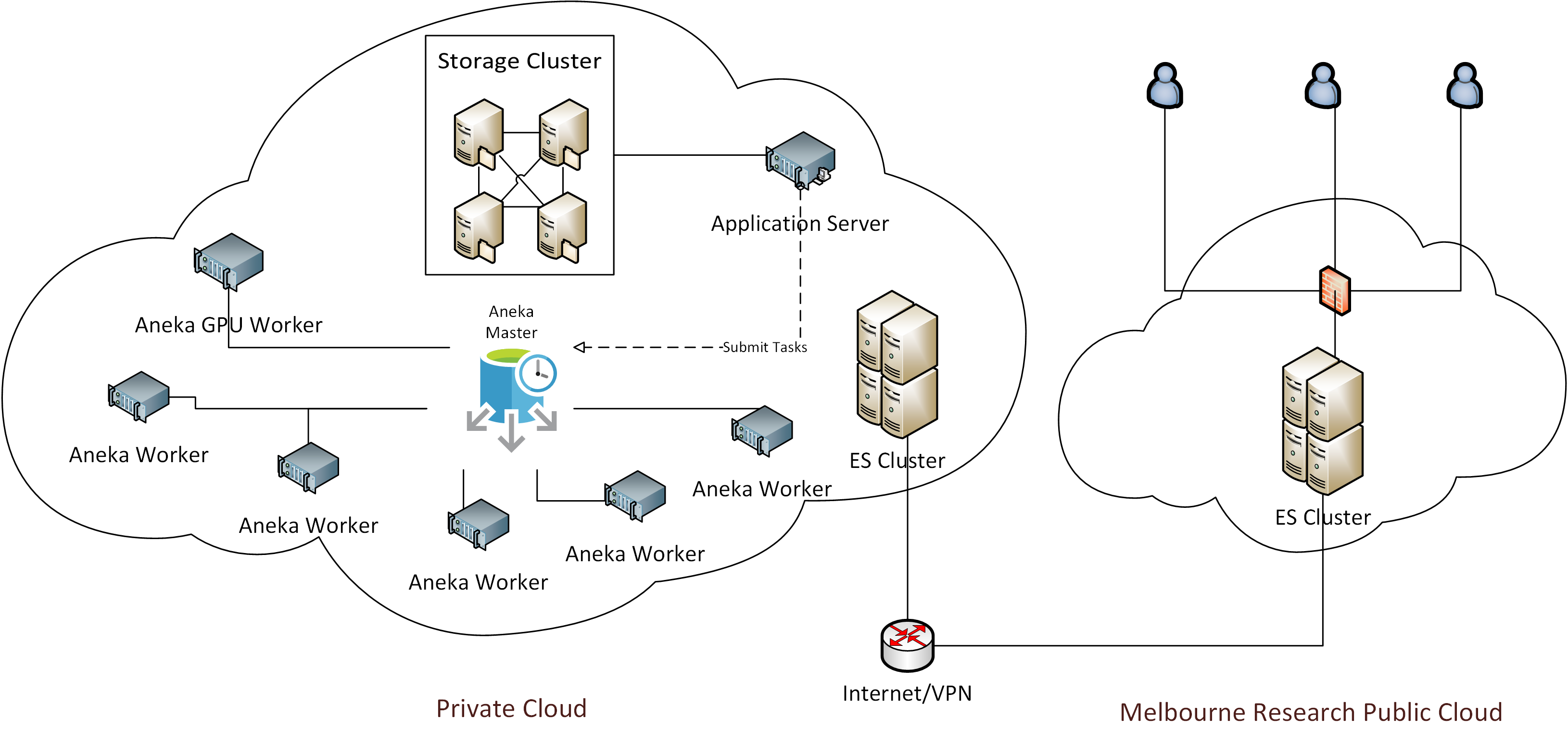}}
	\caption{Architecture Diagram}
	\label{fig:architecture}
\end{figure*}

This section outlines an overview of the system architecture visualized in Fig. \ref{fig:architecture}. The system utilizes a hybrid cloud environment: an on-premise private cloud for storing and processing data and Melbourne Research Cloud (MRC) for hosting and serving public-facing ES traffic.

While designing a data-centric processing system, the first thing to consider is how to store and distribute the data efficiently. Although Aneka PaaS supports data distribution via task payload and FTP, it is not efficient and scalable in our usecase. Therefore, we decide to use a centralized storage cluster powered by Minio\cite{minio}, an open-source object storage software. Minio provides S3 compatible API and shared-nothing architecture for scalability and availability. It is a shared file system that can be accessed by all Aneka workers/master and application server. When requirement arises, data can be easily replicated to external CSP over VPN connection, creating a multi-cloud storage cluster. The second component is the computing service provided by Aneka PaaS as described in Section \ref{sec:background}. Aneka makes it easy to perform parallel processing by encapsulating task scheduling and execution. The third component is an ES cluster replicated to a MRC public instance over the Internet/VPN. The on-premise primary ES cluster is for data persistence. The secondary instance is configured for read-only accessing and serving public-facing traffic. Due to resource constraint, we are running on a single node at both sides with regular automated snapshot/backup.

The architecture design is quite straightforward, but it addresses some common system design principles as follows:
\begin{enumerate}
\item Scalability: All major components are horizontally and vertically scalable. When additional capability is requisite, one can easily scale up or out by providing more powerful nodes or add more nodes.  
\item Availability: Minio and ElasticSearch use shared-nothing architecture, which is designed for highly available systems. Aneka also has robust mechanism to handle task/node failure automatically.
\item Stability: This can be achieved using fail-fast and idempotent processing. If a task fails for any reason, it can be rescheduled either periodically or automatically without affecting the whole system.
\item Performance: High performance is achieved by facilitating Aneka's distributed scheduling and execution capability.
\item Portability: Application developed with the Aneka SDK/API is portable. It can execute in any environment that supports .Net framework, regardless of cloud service provider. Portability can be easily achieved by moving the application to another CSP like AWS, Azure, GCP, etc. 
\end{enumerate}

The main system logic is implemented in the application server component. The application server works as a collaborator performing some housekeeping tasks. It periodically checks for dataset update, downloads PDF files from various sources and submits tasks to Aneka master. It also monitors the ML model bucket. When a new training set becomes available that can improve the ML model, the application server submits a task to Aneka master for training. The actual data processing tasks are performed by worker nodes.

\section{System Implementation}\label{sec:sysimpl}
\begin{figure*}[htbp]
	\centering
	\includegraphics[width=0.9\textwidth]{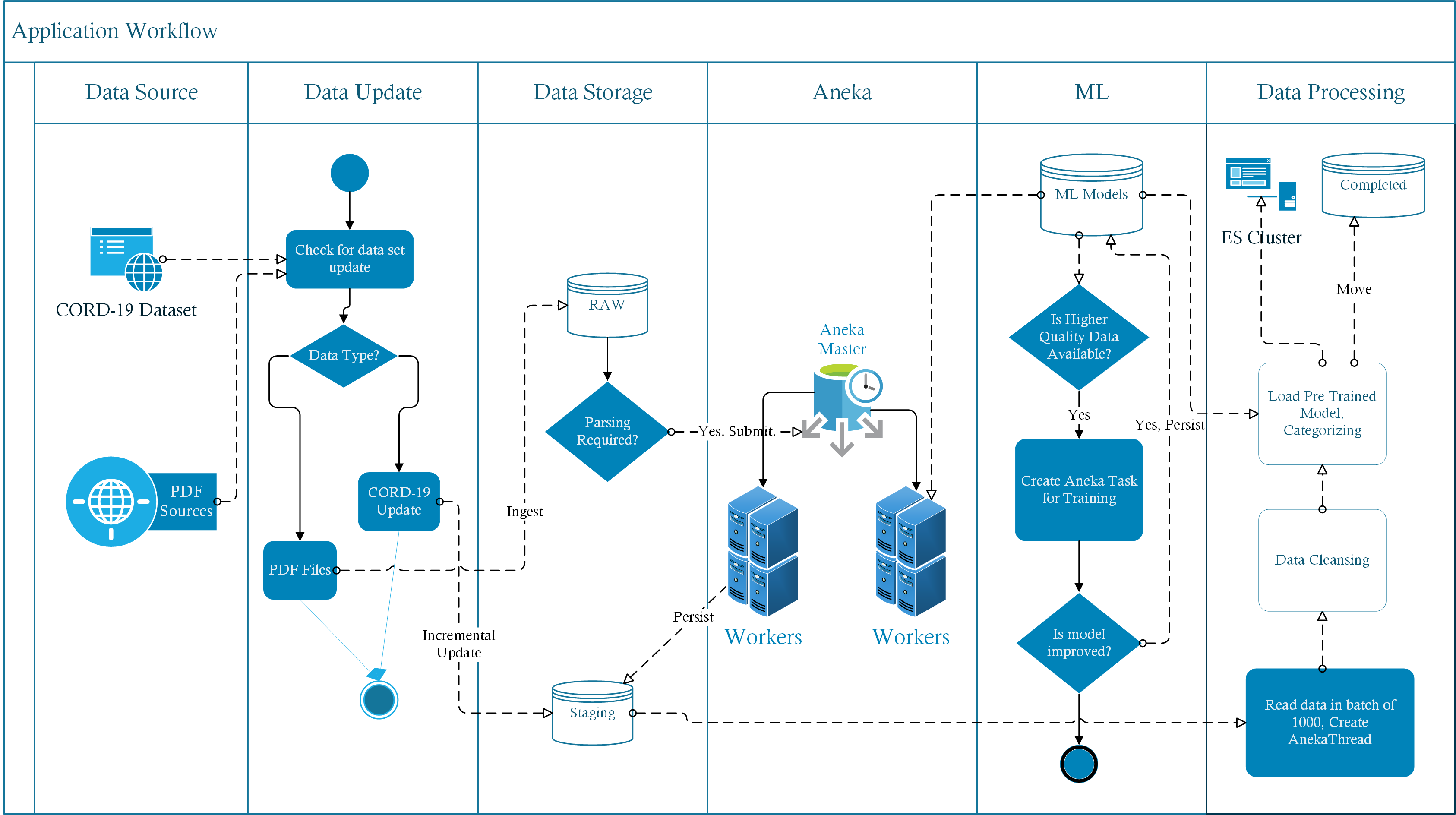}
	\caption{Application Workflow}
	\label{fig:workflow}
\end{figure*}

This section describes system implementation shown in Fig. \ref{fig:workflow}. Firstly, we describe four storage buckets followed by their purposes. Afterwards, the application workflows are explained. 

There are four types of data stored in four buckets:
\begin{enumerate}
	\item RAW: This bucket contains PDF files need to be extracted.
	\item Staging: This bucket contains TEI/XML, metadata and JSON files that have been extracted from PDF files or downloaded from the public dataset.
	\item Completed: Processed files are moved from the staging area to this bucket for archiving/referencing purposes. This also avoids duplicate processing.
	\item ML Models: Dataset for ML Model training and pre-trained model are stored in this bucket.
\end{enumerate}

The first workflow is checking for data source update. The application checks for the latest version from CORD-19 dataset and other sources. If new data is available in CORD-19 dataset, it ingests incremental data into the staging bucket. Since every article is binary checksummed with SHA1 and all processed articles are stored in a separated bucket, an incremental update is possible by comparing checksums. Because checksum values are used as object keys in S3/Minio, complexity for checking is O(1). If new PDF files are available from various sources, they are stored in the RAW bucket first. Periodically, RAW PDF files are sent as AnekaTask for text extracting using Grobid on worker nodes.

Secondly, the ML model bucket is checked periodically. We use SdcaMaximumEntropy multi-class trainer provided by ML.net to train our model and save it to the bucket. Since high quality training dataset is crucial for achieving higher accuracy, if better quality labelled data is available, the system will train a newer model and compare its accuracy with the old one, keeping the better model for later inferencing. 

\begin{algorithm}
	\SetAlgoLined
	\KwResult{Performing ETL and categorizing articles in parallel}
	Aneka Initialisation\;
	totalNumberOfFiles $\leftarrow$ Count number of files in staging bucket\;
	\ForEach{Batch of 1000 files}
	{
		keys[1000] $\leftarrow$ file object keys\;
		Create an AnekaThread wrapping the keys and process logic\;
		Submit to Aneka Master to schedule the processing\;
	}
	$\parallel$ Worker Nodes in parallel:\\
	model $\leftarrow$ Load pre-trained ML model\;
	\ForEach{key in keys}{
			Read file from storage server\;
			Clean unnecessary texts, we only keep metadata, abstract and full-text\;
			Predict the category with model and add labels\;
			Send data to ES cluster\;
			Move the processed file to the completed bucket\;
	}
	\caption{Data Processing Workflow.}
	\label{alg:dataprocessing}
\end{algorithm}

\begin{figure*}[h]
	\centering
	\includegraphics[width=0.9\textwidth]{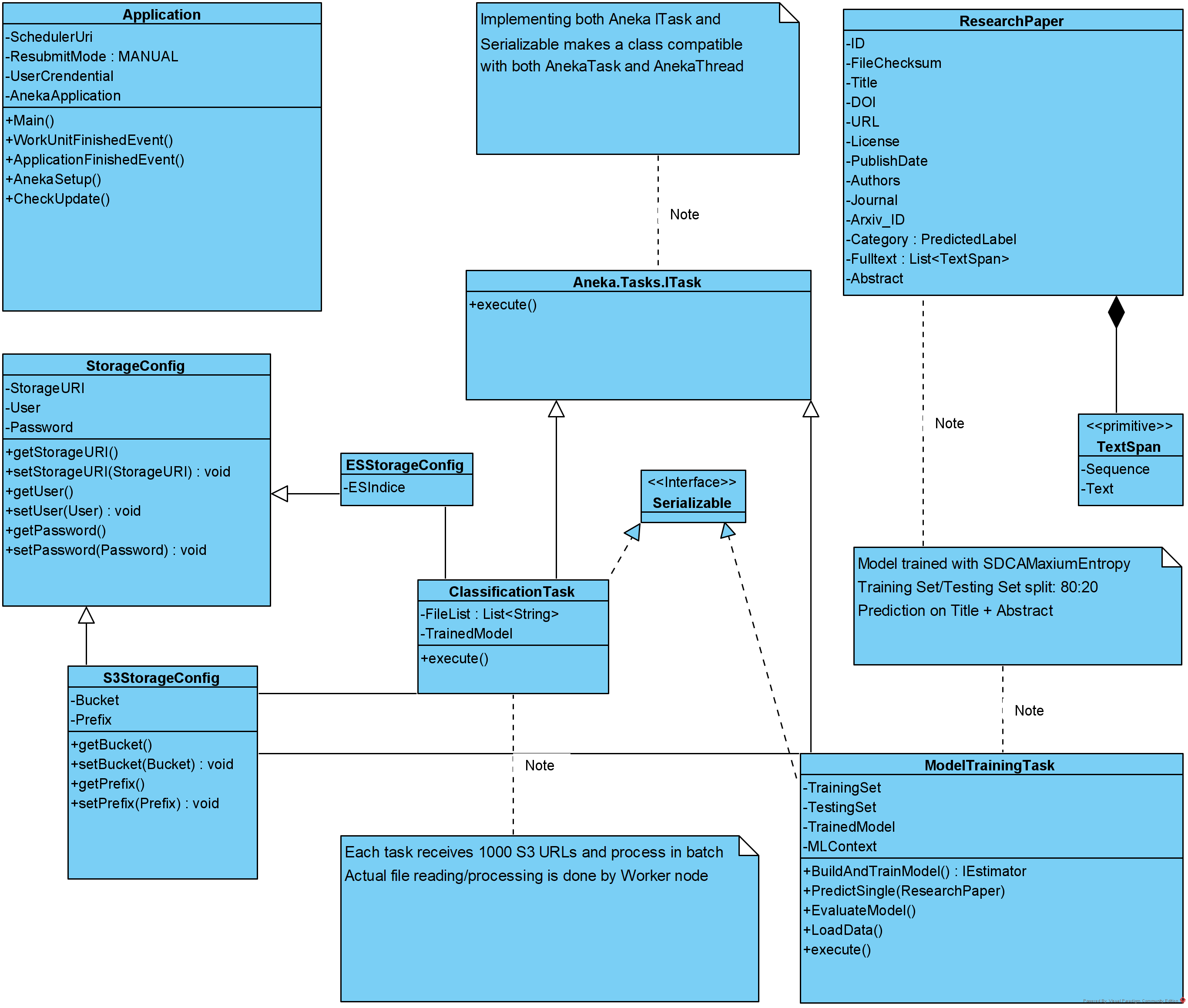}
	\caption{UML Class Diagram}
	\label{fig:uml}
\end{figure*}

The main data processing workflow is described in the following algorithm \ref{alg:dataprocessing}. The UML class diagram in Fig. \ref{fig:uml} shows classes comprised the application. To define an Aneka task, developers only need to make the class sterilizable and implement Aneka.Tasks.ITask interface with only a single method to program the execution logic.

\begin{figure*}[ht]
	\begin{minipage}{\columnwidth}
			\includegraphics[width=\linewidth]{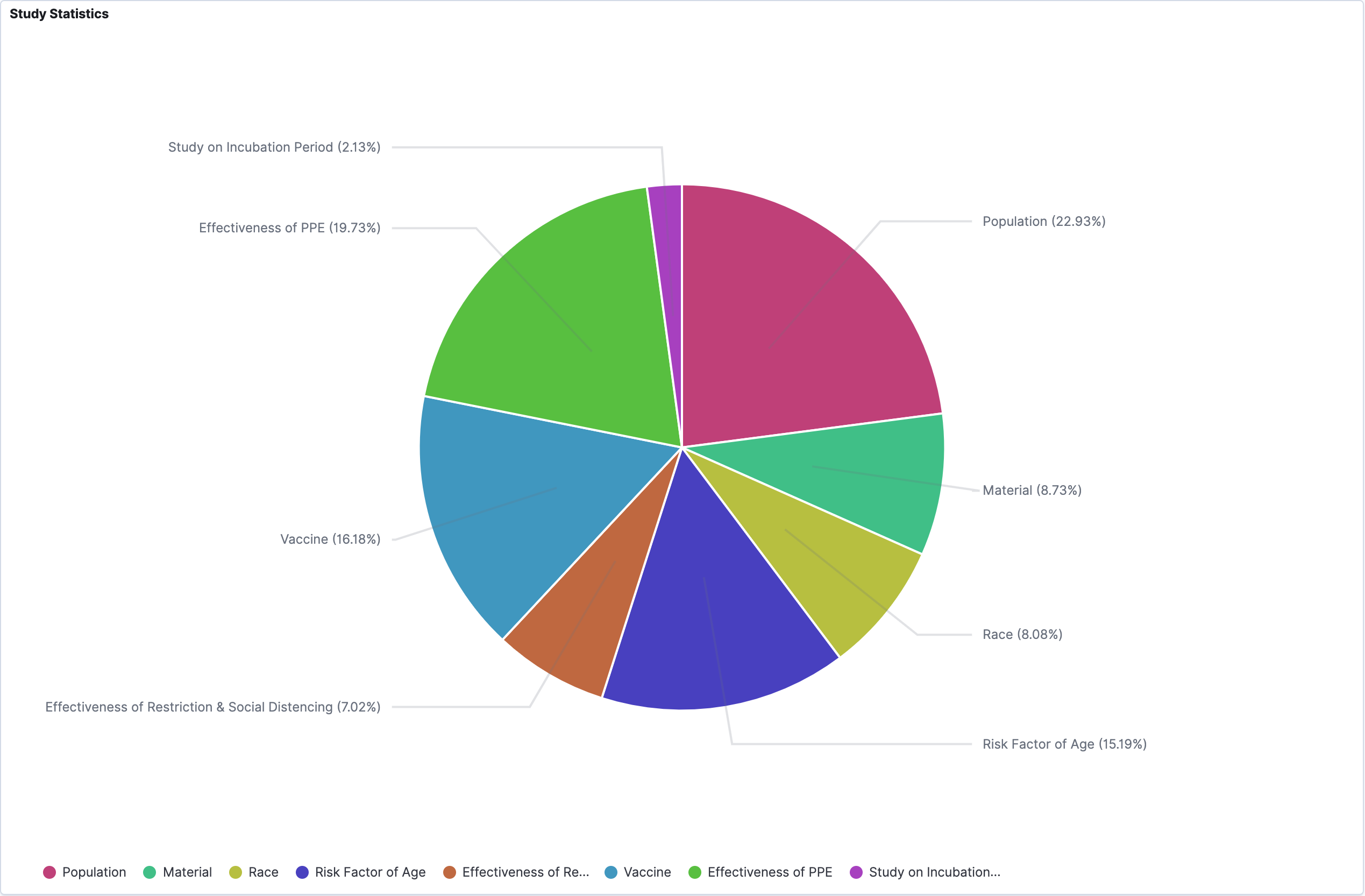}
		\caption{Field of Studies}
		\label{fig:fieldofstudy}
	\end{minipage}
\begin{minipage}{\columnwidth}
	\includegraphics[width=\linewidth]{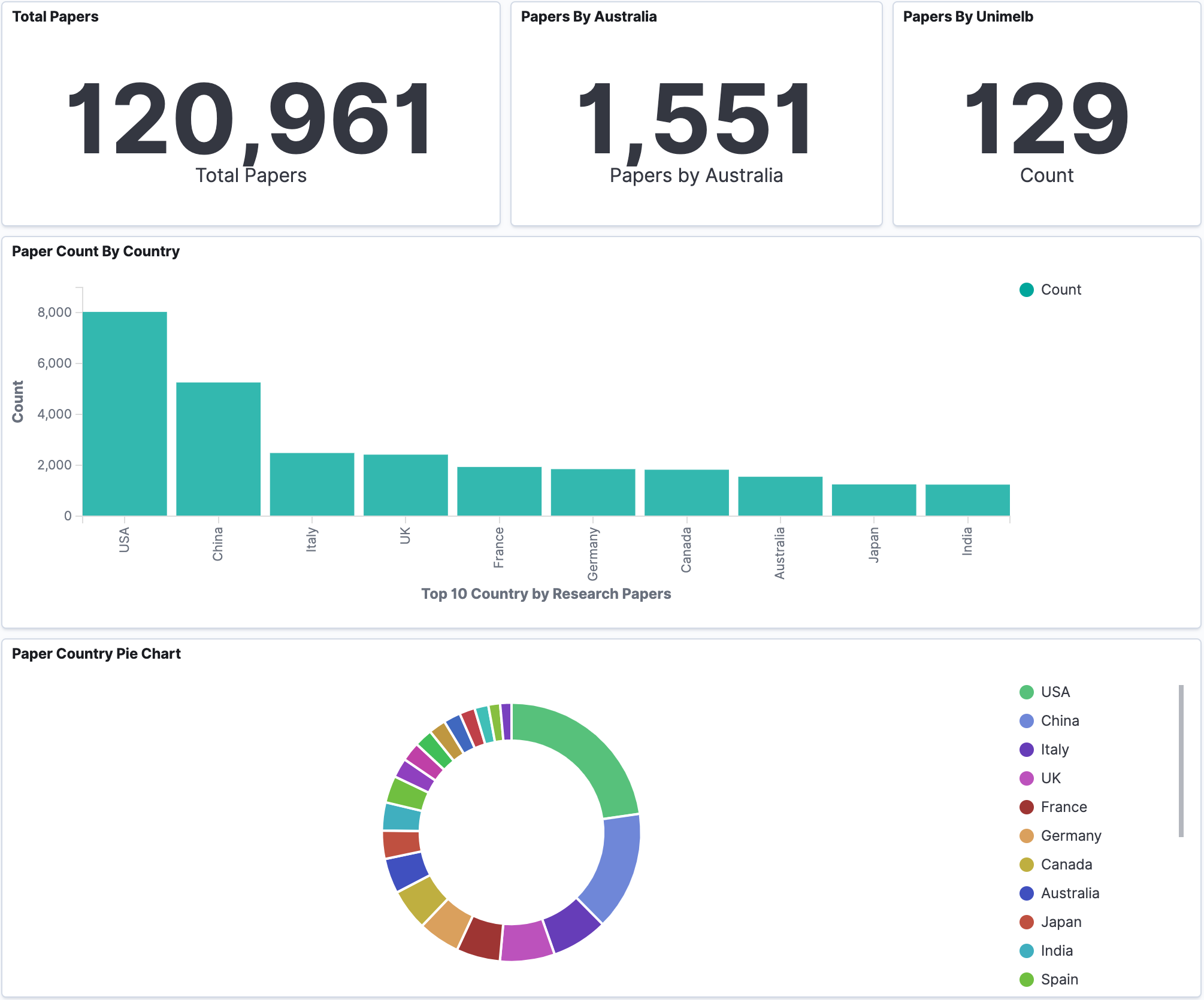}
\caption{Contribution by Country}
\label{fig:contibbycountry}
\end{minipage}
\end{figure*}

Moving on to the application demonstration, Fig. \ref{fig:fieldofstudy} and Fig. \ref{fig:contibbycountry} illustrate two example queries we searched, aggregated and visualized on ElasticSearch and Kibana as general interest. Since the data is ever changing, these are exported at the time of querying and for illustration purposes only.

Question 1: What are the hottest research areas?
The result shows population(virus spread patterns), vaccine, Effectiveness of PPE and risk factors are among the most studied areas.
Question 2: Which country is contributing the most efforts to these researches? How many articles are contributed by Australia and Unimelb?

\section{Performance Evaluation}\label{sec:eval}
In this section, we introduce our testing environment, present benchmark results and also share our observations based on the experimental findings.
\subsection{Testing Environment Setup}
Table \ref{tab:specs} lists the specifications of our test environment. The Aneka master was setup to run in a converged mode, which means it also doubles as a worker node. In our performance benchmark, the single node execution was performed on the master node. In our testbed, the nodes are virtualized instances with Linux KVM running on top of three physical hosts. Each host has 2x1 GBE configured with LACP(802.3ad), layer 2+ hashing and connected to the same gigabit managed switch.
\begin{table}[htbp]
\begin{center}
\begin{tabular}{ |c||c|}
	\hline
	\multicolumn{2}{|c|}{Testing Environment} \\
	\hline
	Role x Qty& CPU/RAM/OS \\
	\hline
	Aneka Master x1& Xeon E5-2690v3 2.6GHz(4C)/8G/Windows 10\\
	Aneka Worker x3& Xeon E5-2690v3 2.6GHz(4C)/8G/Windows 10\\
	Minio x1 & Xeon E3-1245v5 3.5GHz(2C)/8G/Ubuntu 20.04\\
	ES Private x1 & Xeon E3-1245v5 3.5GHz(4C)/16G/Ubuntu 20.04\\
	ES Public x1 & MRC vCPU(2C)/8G/Ubuntu 18.04\\
	\hline
\end{tabular}
\vspace{5px}
\caption{Specification for Testing Environment}
\label{tab:specs}
\end{center}
\end{table}

\subsection{Benchmark Results and Observations}
\begin{figure*}[ht]
\begin{minipage}{\columnwidth}
			\includegraphics[width=\linewidth]{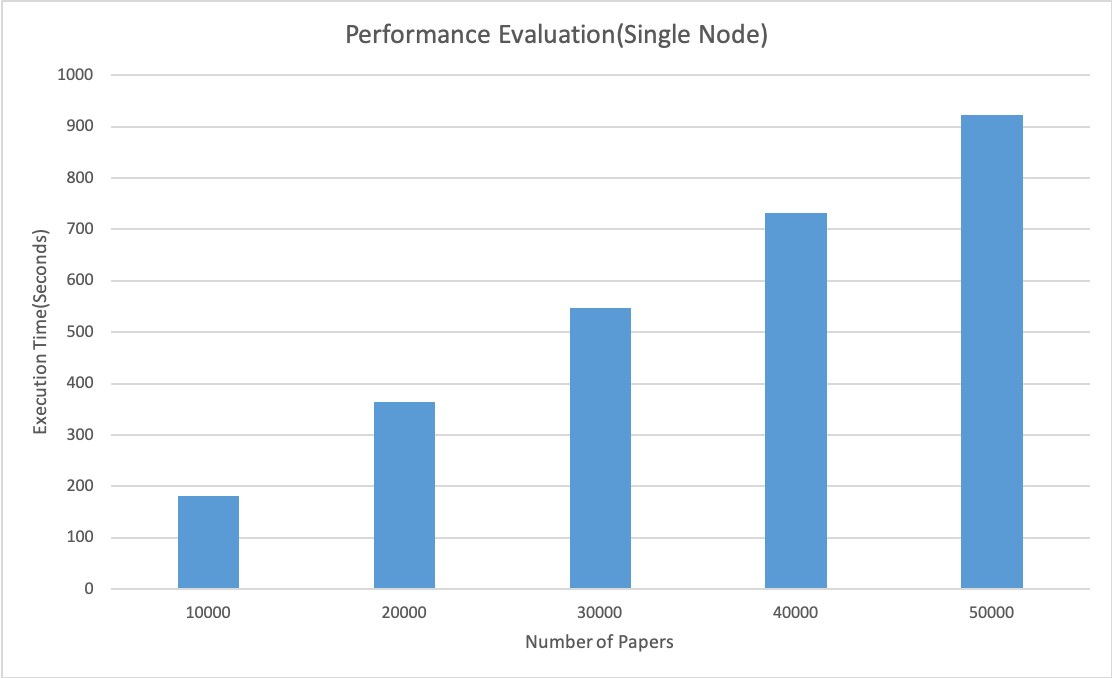}
			\caption{Single Node Benchmark}
	\label{fig:benchsingle}
\end{minipage}
\hfill
\begin{minipage}{\columnwidth}
	\includegraphics[width=\linewidth]{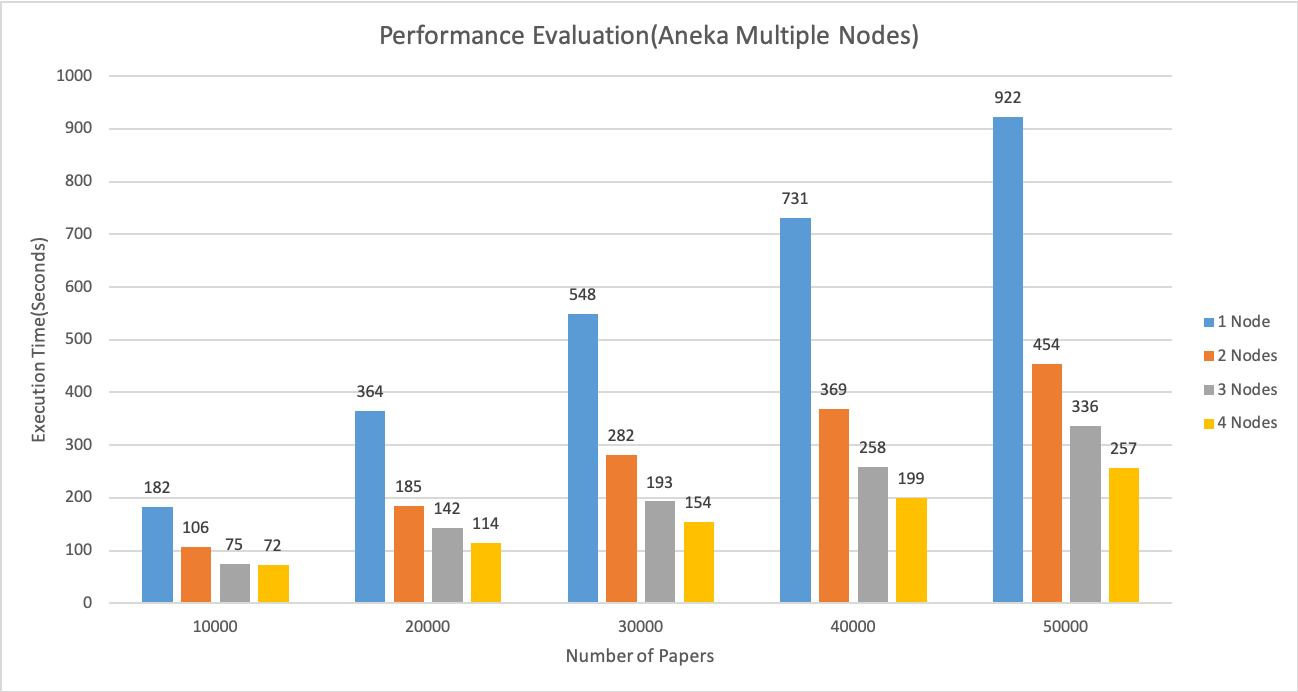}
	\caption{Single vs Multi Node Benchmark}
			\label{fig:benchmulti}
\end{minipage}
\end{figure*}

The experiments were done with smaller sets of articles \\
 $N = \{10000,20000,30000,40000,50000\}$  randomly taken from CORD-19 dataset, running with $M = \{1,2,3,4\} $ Nodes to demonstrate performance and scalability. Fig. \ref{fig:benchsingle} and Fig. \ref{fig:benchmulti} show execution time for a single node and multiple nodes.

As expected, for a single node, the execution time increases linearly in relation to the size of input. For multiple nodes, the processing time can  be significantly reduced; we are able to achieve near linear scalability with little overhead.
There are few observations worth note here:
\begin{enumerate}
\item Processing time of each article is near constant, the fluctuation is relatively small from 18.2ms to 18.7ms. This is largely caused by I/O bound and network latency since each article is read from storage server before processing and persisted to the ES cluster afterwards. 
\item Using two nodes theoretically reduces processing time by half comparing to a single node. The actual result is less than 2x since there is communication and task scheduling overhead.
\item When dataset is not large enough, after reaching the diminish return point, further increasing number of nodes will not reduce processing time much further as theoretically expected due to overheads. e.g. for 10000-20000 articles, the benefit of using more than two nodes is less rewarding, meaning using two nodes is the best in this problem size. For larger input, using more nodes is still beneficial.
\end{enumerate}

\section{Conclusions and Future Work}
In this paper, we proposed a system architecture for indexing and analysing scholarly articles, in particular, the CORD-19 dataset. We also presented an application design and implementation. By using the Aneka PaaS solution, parallel data processing application can be effortlessly developed. It significantly reduces entry barrier for a developer to develop such a distributed system. 

For future work, the ML model can always be improved with higher quality labelled datasets. A significant contribution is the CODA-19 dataset\cite{Huang2020}. The authors used 248 human workers provisioned by AWS Mechanical Turk and created a human-annotated dataset. We may utilize this dataset to improve our model but it requires some additional work.

Also, we are planning to test the system to the limit with a more extensive collection. S2ORC\cite{lo-wang-2020-s2orc} is a general-purpose corpus containing 136M+ paper nodes with 12.7M+ full-text papers(as of 27/July/2020) from many different sources. The future direction will be testing/benchmarking the system design with a much larger dataset and make improvements over iterations.

\section{Acknowledgement}
The public ElasticSearch and Kibana instance  is hosted on Melbourne Research Cloud(MRC). We thank MRC for providing computing resource and bandwidth. 

\bibliographystyle{IEEEtran}
\bibliography{reference.bib}

\begin{thebibliography}{10}
\providecommand{\url}[1]{#1}
\csname url@samestyle\endcsname
\providecommand{\newblock}{\relax}
\providecommand{\bibinfo}[2]{#2}
\providecommand{\BIBentrySTDinterwordspacing}{\spaceskip=0pt\relax}
\providecommand{\BIBentryALTinterwordstretchfactor}{4}
\providecommand{\BIBentryALTinterwordspacing}{\spaceskip=\fontdimen2\font plus
\BIBentryALTinterwordstretchfactor\fontdimen3\font minus
  \fontdimen4\font\relax}
\providecommand{\BIBforeignlanguage}[2]{{%
\expandafter\ifx\csname l@#1\endcsname\relax
\typeout{** WARNING: IEEEtran.bst: No hyphenation pattern has been}%
\typeout{** loaded for the language `#1'. Using the pattern for}%
\typeout{** the default language instead.}%
\else
\language=\csname l@#1\endcsname
\fi
#2}}
\providecommand{\BIBdecl}{\relax}
\BIBdecl

\bibitem{wang-lo-2020-cord19}
\BIBentryALTinterwordspacing
L.~L. Wang, K.~Lo, Y.~Chandrasekhar, R.~Reas, J.~Yang, D.~Eide, K.~Funk,
  R.~Kinney, Z.~Liu, W.~Merrill, P.~Mooney, D.~Murdick, D.~Rishi, J.~Sheehan,
  Z.~Shen, B.~Stilson, A.~D. Wade, K.~Wang, C.~Wilhelm, B.~Xie, D.~Raymond,
  D.~S. Weld, O.~Etzioni, S.~Kohlmeier, D.~Burdick, D.~Eide, K.~Funk,
  Y.~Katsis, R.~Kinney, Y.~Li, Z.~Liu, W.~Merrill, P.~Mooney, D.~Murdick,
  D.~Rishi, J.~Sheehan, Z.~Shen, B.~Stilson, A.~D. Wade, K.~Wang, N.~X.~R.
  Wang, C.~Wilhelm, B.~Xie, D.~Raymond, D.~S. Weld, O.~Etzioni, and
  S.~Kohlmeier, ``{CORD-19: The Covid-19 Open Research Dataset},'' in
  \emph{Proceedings of the Workshop on {\{}NLP{\}} for {\{}COVID-19{\}} at
  {\{}ACL 2020{\}}}.\hskip 1em plus 0.5em minus 0.4em\relax Association for
  Computational Linguistics, Jul 2020. [Online]. Available:
  \url{http://arxiv.org/abs/2004.10706 https://arxiv.org/abs/2004.10706}
\BIBentrySTDinterwordspacing

\bibitem{kaggle}
\BIBentryALTinterwordspacing
``{COVID-19 Open Research Dataset Challenge (CORD-19) | Kaggle}.'' [Online].
  Available:
  \url{https://www.kaggle.com/allen-institute-for-ai/CORD-19-research-challenge/data}
\BIBentrySTDinterwordspacing

\bibitem{Calheiros2012}
\BIBentryALTinterwordspacing
R.~N. Calheiros, C.~Vecchiola, D.~Karunamoorthy, and R.~Buyya, ``{The Aneka
  platform and QoS-driven resource provisioning for elastic applications on
  hybrid Clouds},'' \emph{Future Generation Computer Systems}, vol.~28, no.~6,
  pp. 861--870, 2012. [Online]. Available:
  \url{http://dx.doi.org/10.1016/j.future.2011.07.005}
\BIBentrySTDinterwordspacing

\bibitem{Vecchiola2009}
C.~Vecchiola, X.~Chu, and R.~Buyya, ``{Aneka: A Software Platform for .NET
  Based Cloud Computing},'' \emph{Advances in Parallel Computing}, vol.~18, pp.
  267--295, 2009.

\bibitem{awscord19}
\BIBentryALTinterwordspacing
``{CORD-19 Search}.'' [Online]. Available: \url{https://cord19.aws/{\#}}
\BIBentrySTDinterwordspacing

\bibitem{azurecord19}
\BIBentryALTinterwordspacing
``{COVID-Miner}.'' [Online]. Available:
  \url{https://dain.research.cchmc.org/covidminer/}
\BIBentrySTDinterwordspacing

\bibitem{tekstack}
\BIBentryALTinterwordspacing
``{TEKStack Health - COVID-19 Research Portal}.'' [Online]. Available:
  \url{https://covid-research.tekstackhealth.com/}
\BIBentrySTDinterwordspacing

\bibitem{covidminer}
\BIBentryALTinterwordspacing
``{COVID-Miner}.'' [Online]. Available:
  \url{https://dain.research.cchmc.org/covidminer/}
\BIBentrySTDinterwordspacing

\bibitem{Wolinski2020}
\BIBentryALTinterwordspacing
F.~Wolinski, ``{Visualization of Diseases at Risk in the COVID-19
  Literature},'' May 2020. [Online]. Available:
  \url{http://arxiv.org/abs/2005.00848}
\BIBentrySTDinterwordspacing

\bibitem{covidseer}
\BIBentryALTinterwordspacing
``{COVIDSeer}.'' [Online]. Available:
  \url{https://covidseer.ist.psu.edu/search?query=covid}
\BIBentrySTDinterwordspacing

\bibitem{covidexplorer}
\BIBentryALTinterwordspacing
``{COVID Explorer}.'' [Online]. Available:
  \url{https://coronavirus-ai.psu.edu/database}
\BIBentrySTDinterwordspacing

\bibitem{mldotnet}
J.~McCaffrey, ``{ML.NET: The Machine Learning Framework for .NET Developers.}''
  \emph{MSDN magazine}, no.~13, p.~24, 2018.

\bibitem{grobid}
\BIBentryALTinterwordspacing
``{GROBID}.'' [Online]. Available: \url{https://github.com/kermitt2/grobid}
\BIBentrySTDinterwordspacing

\bibitem{minio}
\BIBentryALTinterwordspacing
``{MinIO | High Performance, Kubernetes Native Object Storage}.'' [Online].
  Available: \url{https://min.io/}
\BIBentrySTDinterwordspacing

\bibitem{elastic}
\BIBentryALTinterwordspacing
``{Elastic Stack: Elasticsearch, Kibana, Beats {\&} Logstash | Elastic}.''
  [Online]. Available: \url{https://www.elastic.co/elastic-stack}
\BIBentrySTDinterwordspacing

\bibitem{NadjaranToosi2018}
A.~{Nadjaran Toosi}, R.~O. Sinnott, and R.~Buyya, ``{Resource provisioning for
  data-intensive applications with deadline constraints on hybrid clouds using
  Aneka},'' \emph{Future Generation Computer Systems}, vol.~79, pp. 765--775,
  Feb 2018.

\bibitem{Huang2020}
\BIBentryALTinterwordspacing
T.-H.~K. Huang, C.-Y. Huang, C.-K.~C. Ding, Y.-C. Hsu, and C.~L. Giles,
  ``{CODA-19: Reliably Annotating Research Aspects on 10,000+ CORD-19 Abstracts
  Using a Non-Expert Crowd},'' May 2020. [Online]. Available:
  \url{http://arxiv.org/abs/2005.02367}
\BIBentrySTDinterwordspacing

\bibitem{lo-wang-2020-s2orc}
\BIBentryALTinterwordspacing
K.~Lo, L.~L. Wang, M.~Neumann, R.~Kinney, and D.~S. Weld, ``{S2ORC: The
  Semantic Scholar Open Research Corpus},'' in \emph{Proceedings of ACL}, Nov
  2019. [Online]. Available: \url{http://arxiv.org/abs/1911.02782}
\BIBentrySTDinterwordspacing

\end{thebibliography}
\end{document}